\documentstyle[aps,prl]{revtex} % used when sending to eprint
\newcommand{\BE}{\begin{equation}}
\newcommand{\EE}{\end{equation}}
\newcommand{\BA}{\begin{eqnarray}}
\newcommand{\EA}{\end{eqnarray}}

\begin{document}
\draft

\twocolumn[\hsize\textwidth\columnwidth\hsize\csname@twocolumnfalse\endcsname
\title{Scaling behavior in wave localization}
\author{Zhen Ye}
\address{Department of Physics, National Central University, Chungli, Taiwan 32054}
\date{April 22, 2001}

\maketitle

\begin{abstract}

Wave localization is a ubiquitous phenomenon. It refers to
situations that transmitted waves in scattering media are {\it
trapped} in space and remain confined in the vicinity of the
initial site until dissipated. Based on a scaling analysis, the
localization behavior in two and three dimensions is studied. It
is shown that the localization transition is possible in two
dimensional systems, supporting the recent numerical results.

\end{abstract}

]

When propagating through a medium with many scatterers, waves will
be repeatedly scattered to establish a process of multiple
scattering of waves. It is now well known that multiple scattering
gives rise to many fascinating phenomena, including the photonic
or sonic band gaps in periodic structures\cite{ref}, random
lasers\cite{laser}, and electrical resistivity\cite{Lee}. An
excellent account of multiple scattering was given in
\cite{Akira}.

Under proper conditions, multiple scattering leads to the unusual
phenomenon of wave localization, a concept introduced by
Anderson\cite{Anderson} to explain the metal-insulator transition
induced by disorders in electronic systems and recently reviewed
by Imada et al.\cite{Imada}. That is, the electronic movement can
be completely stopped due to multiple scattering by a sufficient
amount of impurities in solids. It is believed that once the
electronic movement is stopped, the electrons are trapped in
space. The fact that this effect due to the wave nature of
electrons has led to the conjecture that similar phenomena may
also exist in the propagation of classical waves in randomly
scattering media.

Considerable efforts have been devoted to the investigation of
classical wave localization in random media. In most previous
experimental studies, the apparatus is set up in such a way that
waves are transmitted at one end of scattering sample, then the
scattered waves are recorded either on the other end to measure
the transmission or are received at the transmitting site to
measure the reflection from the sample. In either case, the
measurement was done when both the transmitter and receiver are
located outside the sample. The results are subsequently compared
with the theory developed for classical wave
localization\cite{gang4,Phil} to infer the possible localization
effect. In this way, observations of wave localization effects
have been reported for water wave localization by random
underwater topography\cite{water}, for acoustic waves
\cite{acoustic}, microwaves\cite{microwave1,microwave2}, and
arguably for light\cite{light1,light2,light3,light4}.

In these measurements, two phenomena are thought as the indicator
of localization effects. The first is the enhanced backscattering.
As much discussed in the literature (e.~g.
Refs.~\cite{Lee,Akira,light1}), the wave received at any spatial
point is contributed by wave propagated along various paths. For
the bistatic case, the random scattering leads to a destructive
interference of scattered waves, thus reducing the transmission.
For the backscattering situation, however, any random scattering
path that returns to the transmitting source can always be
followed by two opposite directions. The waves which propagate in
the two opposite directions along a loop will acquire the same
phase and therefore interfere constructively at the transmitting
site, yielding the phenomenon of the enhanced backscattering. The
second indicator is associated with the relation between the wave
transmission and the sample size. The theory\cite{Phil} predicts
that once localization occurs, the wave transmission is expected
to undergo a transition from a linear to quadratic decreasing, and
eventually to follow an exponential decay. In line with the
theory\cite{gang4}, it has been the prevailing view that all waves
are localized in two dimensional (2D) systems with any amount of
disorders.

The recent numerical simulation, however, shows that waves are not
always localized in 2D random systems\cite{PRL,APL}. The work is
done with acoustic propagation in water containing many randomly
placed air-filled cylinders, by an exact method for multiple
scattering. Unlike most previous cases, the numerical simulation
has been done by placing an acoustic source inside the random
array. The acoustic transmission for various frequencies is
recorded by a receiver located outside the scattering array. It is
found that while in a range of frequencies and for a sufficient
amount of the air cylinders, the transmitted acoustic waves are
indeed trapped or localized inside the random medium, the waves
remain extended outside the localized regime.

An immediate criticism on this observation may be that the
apparent state of wave propagation is an artifact of the finite
size effect. In other words, it may be argued that the
localization length exceeds the size of the scattering medium,
thus the transmitted waves appear to be non-localized. Although it
is true that it is impossible to simulate an infinite scattering
medium, the observed phase transition between localized and
extended wave transmission is not caused by the finite size
effect\cite{PRL}. Two main reasons support this viewpoint. One,
according to the analytic results\cite{local}, the localization
length scales as $L\exp(L/\lambda)$, where $L$ is the mean free
path of the scattering and $\lambda$ is the wavelength. If waves
were localized for all frequencies, the theoretical calculation
would lead to a smooth variation in the localization length. This
is not observed. Second, as pointed out in \cite{APL}, the wave
localization would necessarily lead to a phase ordering. This can
be understood as follows. In terms of wave field, $u$, the energy
flow may be calculated from $\vec{J(\vec{r}}) \sim
i\mbox{Re}[u(\vec{r})\nabla u(\vec{r})]$. Writing
$u=|u|e^{i\theta}$, the flow becomes $\vec{J} \sim
|u|^2\nabla\theta$. Obviously, the energy will be localized when
phase $\theta$ is constant (coherence) and $|u|\neq 0$. Vice
versa, the vanishing energy flow should lead to the phase
ordering. In the simulation in \cite{PRL}, as the sample size is
enlarged, there is no tendency in the phase coherence for
frequencies outside the localization regime. In spite of these,
the fact that the results are numerical is discomforting.

{\small
\input{epsf}
\begin{figure}[hbt]
\begin{center}
\epsfxsize=1.5in \epsffile{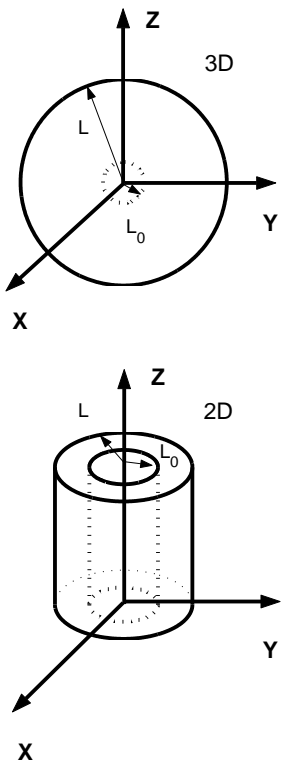} \vspace{12pt} \caption{
\label{fig1}\small Conceptual layout for 2D and 3D systems.}
\end{center}
\end{figure}
}

We are therefore naturally led to the question of how to reconcile
the contradiction between the numerical observation in \cite{PRL}
and the previous assertion that all waves are localized in 2D
disordered systems. In this Letter, we wish to present an
examination of the problem. In particular, following the method of
the scaling analysis\cite{gang4} we will show that wave trapping
behavior in 2D bears similarities to that in three dimension (3D).
Under appropriate conditions, the localization transition, i.~e.
the transition from the extended or propagating state to the
localized state as usually seen in 3D, is not impossible in 2D,
thus providing an explanation to the observation in \cite{PRL}.

Before moving on, we point out a few interesting properties
associated with the acoustic scattering by parallel air-cylinders
in water. (1) The air-cylinders in water are strong acoustic
scatter due to the large contrast in the acoustic impedance. At
low frequencies, there appears a resonant scattering. (2) In a
wide range of frequencies above the resonance, the scattering is
nearly isotropic, in the way that the backward scattering strength
is not negligible compared to the forward scattered strength. In
most previous cases, the backscattered wave is neglected. This
approximation is only valid for weak tenuous
scattering\cite{weak}. (3) Though complicated, the scattering by
an array of many air-cylinders allows for an exact formulation and
can be evaluated to desired degrees of accuracy\cite{PRL,Twersky}.
And it is fair to mention that the formulation has been applied
successfully to inspect the recent experiments\cite{Chen}. (4)
Experimentally, the air-filled cylinders can be any gas enclosure
with a thin insignificant elastic shell. In short, these
properties make air-cylinders in water an ideal 2D system for
theoretical and experimental localization studies.

Now we analyze the acoustic localization in water with many
air-cylinders. Upon inspection, we believe that the evident
contradiction with the previous claim is due to the difference in
the ways that the wave localization is inferred or interpreted. It
has been thought that enhanced backscattering is a precursor to
localization. Our numerical results shows, however, that there is
no direct link between backscattering enhancement and
localization\cite{AAD}. Furthermore, as mentioned above, most
previous measurements are performed when both transmitter and
receiver are located outside the scattering medium. The study of
whether the waves are localized or extended is obscured by
boundary effects such the reflection and deflection effects. These
effects attenuate waves, resulting in possibly an exponential
decay in transmission and thus making the data interpretation
ambiguous. It is highly plausible that the inhibition in the wave
transmission does not necessarily guarantee that the wave can be
actually trapped in the medium once the transmitting source is
moved into the randomly scattering medium. In other words, it is
necessary to differentiate the situation that the wave is blocked
from transmission from the situation that wave can be actually
localized in the medium; we believe that the latter case is in
fact what the concept of wave localization is meant to
be\cite{dict}. We stress that whether waves are localized or
extended is an intrinsic property of the system that is supposed
to be infinite. This property does not depend on the source, and
should not depend on the boundary either; thus a genuine analysis
should not be plagued by boundary effects not only in the
localization region but also in the non-localization region. We
believe that while the source is placed inside the medium with
increasing sizes, the infinite system can be mimicked and the
localization property can be probed without ambiguity.

To discern the observation in \cite{PRL}, we adopt a scaling
analysis by analogy with that presented in \cite{gang4}. Consider
that a transmitting source is inside a homogeneous random medium.
To account for the fact that the source is inside the medium, we
take the geometry as shown in Fig.~1. We consider the cylindrical
and spherical scaling for 2D and 3D respectively. In line with the
discussion in \cite{gang4}, for small resistance $R$ the medium is
assumed to follow the ohmic behavior. This leads to \BE R \sim
\left\{\begin{array}{ll} L/L_0, & \mbox{for} \ 1D\\\ln(L/L_0), &
\mbox{for}\ 2D\\ \frac{1}{L_0} - \frac{1}{L}, & \mbox{for} \
3D.\end{array}\right., \label{eq:1} \EE where $L_0$ refers to the
microscopic size\cite{gang4}. This is valid when $R$ is small. In
the other limit that the resistance is large, exponential wave
localization is expected. By taking into the geometric factors,
the resistance thus grows as \BE R\sim L^{d-1}e^{L/L_1},
\label{eq:2}\EE where $d$ denotes the dimension and $L_1$ is the
localization length.

The scaling function is defined as \BE \beta(R(L)) = \frac{d\ln
R}{d\ln L}. \label{eq:3}\EE Taking into account Eqs.~(\ref{eq:1})
and (\ref{eq:2}), the asymptotic behavior for $\beta$ in one
dimension is \BE \beta \sim \left\{
\begin{array}{ll} 1, & \mbox{for} \ \ln(R)\rightarrow -\infty\\
\ln(R), & \mbox{for} \ \ln(R)\rightarrow \infty\end{array}\right.
\label{eq:4}\EE It is clear from this equation, the localization
behavior is the same as that predicted in \cite{gang4}. We can
also obtain the asymptotic behavior for the scaling function in
both 2D or 3D as \BE \beta \sim \left\{\begin{array}{ll}
e^{-\ln(R)}, & \mbox{for}\ \ln(R)\rightarrow -\infty\\ \ln(R), &
\mbox{for} \ \ln(R)\rightarrow
\infty\end{array}\right.\label{eq:5}\EE This equation indicates
that the wave localization behavior in 2D and 3D should be
similar. What is expected for 3D may also appear in 2D.

Equation (\ref{eq:5}) is the basis for our discussion. From the
asymptotic behavior in Eqs.~(\ref{eq:4}) and (\ref{eq:5}), we may
sketch the universal curves in $d=1,2,3$ dimensions. The central
assumption here is continuity\cite{gang4}: once wave is localized,
the increasing sample size would always mean more localization.
This assumption has been discussed in some detail in \cite{gang4}.
It is obvious that the 1D situation is a replicate of that shown
in \cite{gang4}. The result is that all waves are localized in one
dimension for any given amount of disorders.

{\small
\input{epsf}
\begin{figure}[hbt]
\begin{center}
\epsfxsize=2in \epsffile{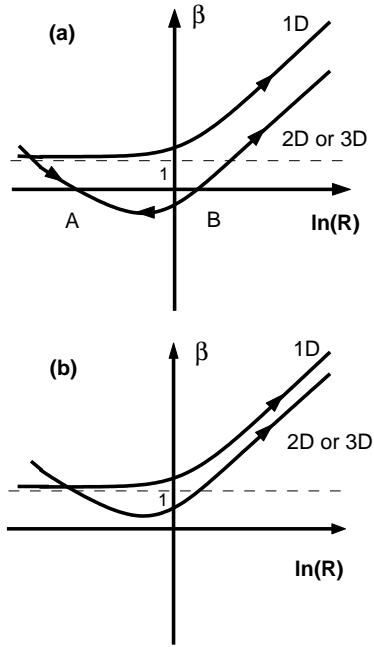} \vspace{12pt} \caption{
\label{fig2}\small Plots of $\beta$ vs $\ln(R)$ for 1, 2, and 3
dimensions.}
\end{center}
\end{figure}
}

The situations in two or three dimensions are more subtle. Two
possibilities are shown in Fig.~2. In the first instance shown by
Fig.~2(a) , as $\ln(R)$ increases, the scaling function $\beta$
may decrease, then crosses the horizontal axis and reaches a
minimum before increasing to follow the linear relation for large
$\ln(R)$. The crossing of the horizontal axis produces two fixed
points: A and B. At both points, $\beta$ vanishes. It is clear
that A and B are respectively the stable and unstable fixed
points. Point B separates the localization state and the extended
state. When $\ln(R)$ is greater than $B$, the increasing sample
size leads to an infinite resistance, thus the waves become
localized {\it inside} the medium. When $\ln(R)$ is below point B,
increasing sample size leads the system to the fixed point A, at
which the increasing $L$ will no longer affect the resistance. On
the first sight, this feature seems awkward. After inspection, it
becomes clear that it is actually a clear indicator of a wave
propagating state, i.~e. the extended state. This can be
understood as follows. As the transmitted wave propagates, the
wave coherence starts to decrease, yielding the way to
incoherence. The total wave is the addition of the coherence and
incoherence waves\cite{Akira}. When there is no absorption, by
energy conservation the total wave transmission, an appropriation
of the inverse of the resistance, will not change along the
propagation path. The transmission will thus not vary as the
sample size changes. Therefore the feature at point A actually
reflects the law of energy conservation. This picture has indeed
been supported by the previous simulations\cite{PRL,AAD}.

The second possibility is shown in Fig.~2(b). The scaling function
$\beta$ will not drop below the horizontal axis. In this case, all
waves in two and three dimensions are localized like in the 1D
situation. For example, this is expected to occur when the amount
of disorders is exceedingly large\cite{PRL,AAD2}. Previous results
affirming that all waves are localized in 2D random media may fit
in this possibility.

Based upon the above scaling analysis, we argue that the
observation of wave localization in 2D reported in \cite{PRL}
follows the behavior illustrated by Fig.~2(a). There are critical
points separating the localized state from the extended state.
When waves are localized in the medium, the waves follow the
exponential localization, as clearly shown by Fig.~3 in
\cite{PRL}. Outside the localized regime, the waves remain
extended in space. The averaged transmission consequently is
nearly constant along the traveling path in the radial direction.
Therefore the observation of \cite{PRL} finds the explanation.

In summary, we have shown a scaling analysis of wave localization
in randomly scattering media. We pointed out that the
differentiation should be made with respect to whether the
transmitting source is placed inside or outside the random media.
The problem with the latter is that the effects from other sources
such as boundary reflection, scattering into other directions
cannot be excluded. These effects may result in phenomena which
could have been attributed incorrectly to the localization effect.
In the analysis when the source is outside the medium\cite{gang4},
the asymptotic behavior in the Ohmic region was derived under the
assumption that the current flows uniformly in one direction. This
is only possible with properly scaled sources and the presence of
confining boundaries, obviously in conflict with the proclamation
that whether it is the localization or extended state is the
intrinsic property of the medium and should not rely on a boundary
nor the source.
%Therefore the analysis of \cite{gang4} is vague.
The vagueness is avoided when the source is inside the medium. The
present analysis shows that wave localizations in 2D and 3D random
systems are similar. The transition from the propagating state to
the localized state is possible in both two and three dimensions.
Finally, we note that recent experiments on electronic systems
also suggest a metal-insulator transition in two dimensions in
contrast to the previous assertion; the mechanism in these systems
is still unclear\cite{EA}.

This work received support from the National Science Council. The
stimulating discussions with Profs. Zhao-Qing Zhang (HKUST) and T.
K. Lee (Academia Sinica) are greatly appreciated, not suggesting
whether they agree or disagree to the paper.

\end{document}